# Coarse-Grained Configurational Polymer Fingerprints for Property Prediction using Machine Learning

Ishan Kumar, Prateek K. Jha*

*Department of Chemical Engineering, IIT Roorkee, Roorkee, Uttarakhand, 247667, India*

**ABSTRACT**

In this work, we present a method to generate a configurational level fingerprint for polymers using the Bead-Spring-Model. Unlike some of the previous fingerprinting approaches that employ monomer-level information where atomistic descriptors are computed using quantum chemistry calculations, this approach incorporates configurational information from a coarse-grained model of a long polymer chain. The proposed approach may be advantageous for the study of behavior resulting from large molecular weights. To create this fingerprint, we make use of two kinds of descriptors. First, we calculate certain geometric descriptors like $R_e^2$, $R_g^2$ etc. and label them as Calculated Descriptors. Second, we generate a set of data-driven descriptors using an unsupervised autoencoder model and call them Learnt Descriptors. Using a combination of both of them, we are able to learn mappings from the structure to various properties of the polymer chain by training ML models. We test our fingerprint to predict the probability of occurrence of a configuration at equilibrium, which is approximated by a simple linear relationship between the instantaneous internal energy and equilibrium average internal energy.

## 1. Introduction

Polymers have a broad range of applications in chemical, biomedical, and material sciences.[1–6] Despite having a whole range of interesting physical properties that can be fine tuned by variations in the polymer molecular weight, composition, and functionalization, the actual design of polymeric systems is quite challenging due to their vast design space. Prevailing approaches of the design of polymeric systems are mostly experimental and experience-driven, and there is a tremendous scope for theoretically/computationally derived design rules that can accelerate the polymer design process. With the development of computational methods and Machine Learning (ML) in recent years, there has been a growth in the number of material databases and hence computational design has also become a feasible and reliable option. Recent works in ML on polymers show that it can serve as a promising tool in both property prediction and molecular design.[7]

In order to develop ML model for polymers, they need to be represented as a vector, which is known as their molecular fingerprint. The most common type of fingerprint is a series of binary digits that represent the presence or absence of particular substructures in the molecule.[7–9] A polymer structure is composed of many simpler chemical units, called monomers or repeat units, which are bonded together to form a long chain macromolecule. Most existing methods use the repeat unit's fingerprint as the fingerprint of the polymer or a weighted-average of repeat units in case of a heteropolymer. This results in a total vector that is invariant to the arrangement of monomer components. Although this strategy is acceptable for short oligomers, it fails for longer polymers where the effects of the arrangement of monomers and polymer configurations cannot be neglected. While some properties of polymers like the density or glass transition temperature depends mostly on chemical details at monomer scale, others (e.g., variation of melt viscosity with molecular weight[10]) are predominantly controlled by the configurational statistics.

*Corresponding Author
 *Email address:* prateek.jha@ch.iitr.ac.in (P.K. Jha)
 ORCID(s): 0000-0001-9844-2875 (P.K. Jha)

In recent years, ML has been extensively used in the design of polymers for a range of desired applications. In doing so, one often has to calculate a certain Quantitative Structure-Activity Relationship (QSAR) value for the polymer and it should be within an appropriate range so that the polymer can be suitably used for that specific application. For instance, in ref. [2], authors propose an ML–assisted molecular design and efficiency prediction for high-performance organic photo-voltaic materials. They use predicted filler factor (FF), short circuit current density ($J_{sc}$), and the open circuit voltage ($V_{oc}$) as the desired QSAR target properties. As another example, a promising mechanism to enhance the efficient phosphorescent OLED materials is thermally activated delayed fluorescence (TADF), which may be studied if the delayed fluorescent rate constant $K_{TADF}$ is selected as the desired property.[4] The general approach that is followed in these studies is that first the structure of the polymer is represented using a Simplified Molecular-Input Line Entry System (SMILES) form. This SMILES string is subsequently converted to a fixed length vector like MACCS (166 bits), Pubchem (876 Bits), FP2 (1020 Bits), Extended Hybridization (1021 Bits), Daylight (1024 Bits), or Morgan (2048 Bits).[2] The training dataset consisting of the polymers and their ground truth labels for the desired QSAR properties is either used from existing databases or created using first-principle calculations. This dataset is used to learn a mapping between structure and property using a regression model like Kernel Ridge Regression or a Neural Network. Upon learning the structure–property relation, we can accept or reject a given polymer based on its predicted value. Further, if the task involves generation of new materials, inverse design of polymers with the required target properties through genetic algorithm or other search algorithms can be performed. The fingerprint plays a vary important role in both these steps and hence improvement in the same would result in better predictive and generative capabilities.

One of the most promising fingerprinting approach of polymers is that adopted in "Polymer Genome" [11] in which a custom fingerprint is created using a hierarchical approach combining three levels of descriptors. The first level of descriptors are atomic level descriptors in which the occurrence of a fixed set of atomic fragments (or motifs) are tracked. The next level used is Quantitative Structure-Property Relationships (QSPR) Descriptors, like molecular quantum numbers, fraction of $sp^3$ carbons, molecular Surface area, van der Waals surface, fraction of aromatic rings, etc. The final level consists of morphological descriptors, like the shortest topological distance between rings, fraction of atoms that are part of side-chains, and the length of the largest side-chain. Properties such as the glass transition temperature strongly depend on such features, which influence the way the chains are packed in the polymer. Polymer Genome shows good results across many QSAR tasks like bandgap, dielectric constant, atomisation energy, glass transition temperature etc. For instance, in ref. [12], authors predict the glass transition temperature of a polymer using 5 different models created through cross validation. The $R^2$ varies from 0.713 to 0.759. In ref. [13] also, authors predict the glass transition of amorphous polymers using MD simulations. They obtain a $R^2$ value of 0.84 using cross validation by using a set of 7 descriptors. These results are less promising than the ones in Polymer Genome[11], where an $R^2$ of 0.944 has been reported. The improvement in results can be partially attributed to the fact Polymer Genome also employs a few configurational level descriptors in the top hierarchy (like length of side chain, distance between chains etc.). In contrast, the entire configuration-level information is missed in the other works as their fingerprint is only atomistic (based on QSAR values of the monomers). Further, we see from Polymer Genome that a combined fingerprint works better across tasks and not just on a specific task, and we believe this will also be the case for long chain polymers.

In this work, we aim to introduce configuration-level information in the fingerprint and test its predictive capability through ML tasks. Since atomistic simulations of long polymer chains are computationally intensive, we employ coarse-grained (CG) Monte Carlo (MC) simulations to develop an ML model. CG modelling approaches approximates the representation of complex structures by lumping several atoms and bonds in the atomistic model into CG units. CG resolution may be defined as the number of CG units divided by the number of atoms in the original atomistic model. A decrease in CG resolution results in reduced computational cost of the CG model but lesser information about the original atomistic model is preserved. One may therefore perform CG simulations at an appropriate CG resolution to capture the physical properties of interest. Here, we use a simple bead-spring model[14], in which the polymer chain is composed of $N$ beads ($j = 0, 1, ..., N − 1$) connected by $N − 1$ massless, harmonic springs ($j = 1, 2, ..., N − 1$). Each bead may represent a single repeat unit or a sub-chain containing multiple repeat units. Chain is considered ideal, i.e., non-bonded interactions between the beads are not considered. The internal energy is calculated as the sum of the individual internal energies of all the springs,

$$U = \sum_{j=1}^{N-1} \frac{1}{2} k_j (b_j - b_{0j})^2 \qquad (1)$$

where $b_j$ is the $b_{0j}$ is the equilibrium length of spring $j$ and $k_j$ is the spring constant of spring $j$, connecting beads $j − 1$ and $j$.

In the next section, we discuss the approach followed in our study, beginning with a description of calculated and learnt descriptors followed by the details of prediction model and MC simulations. This is followed by a discussion of the results of this study.

## 2. Approach

### 2.1. Calculated and Learnt Descriptors

In our proposed fingerprint for long chain polymers, we make use of two kinds of descriptors. First, we use calculated descriptors that are geometric descriptor values, which quantify aspects about the configuration of the polymer and can be used as features in a prediction model. We calculate these values for our dataset and store them alongwith the positions so that they can also be used as part of the fingerprint. This way, we incorporate domain knowledge into our ML model and improve explainability of the prediction models. The calculated descriptors used in this work are the following:

1. The end-to-end distance vector $\vec{R}_e$ is defined as the distance between the two ends of the chain ($\vec{r}_0$ and $\vec{r}_N$). The mean squared end-to-end distance is defined as

$$R_e^2 = \langle (\vec{r}_N - \vec{r}_0) \cdot (\vec{r}_N - \vec{r}_0) \rangle \qquad (2)$$

2. The Radius of Gyration $R_g$ is defined as the root mean squared average distance between the bead positions $\vec{r}_i$ and the center of mass $\vec{r}_{CM}$. It characterises the space a polymer occupies in a given configuration. Squared radius of gyration is defined as

$$R_g^2 = \frac{1}{N} \sum_{i=0}^{N-1} \langle (\vec{r}_i - \vec{r}_{CM}) \cdot (\vec{r}_i - \vec{r}_{CM}) \rangle \qquad (3)$$

3. The 3D Radius of Gyration gives a better intuition of the shape of the polymer in 3-dimensions. The gyration tensor is a tensor that describes the second moments of position of the beads. Instead of using the whole matrix, we use the 3 eigenvalues ($\lambda_x^2, \lambda_y^2, \lambda_z^2$) which indicate the principal axis of the imaginary ellipsoid enclosing the polymer.

$$\overset{\leftrightarrow}{Q} = \begin{bmatrix} Q_{xx} & Q_{xy} & Q_{xz} \\ Q_{yx} & Q_{yy} & Q_{yz} \\ Q_{zx} & Q_{zy} & Q_{zz} \end{bmatrix} \qquad (4)$$

Configurational Polymer Fingerprints for Machine Learningwhere

$$Q_{mn} = \frac{1}{N} \sum_{i=0}^{N-1} \left\langle \left(\vec{r}_i^{(m)} - \vec{r}_{CM}^{(m)}\right) \cdot \left(\vec{r}_i^{(n)} - \vec{r}_{CM}^{(n)}\right) \right\rangle \quad (5)$$

and the superscript indicate the components $(x, y, z)$ of the vector.

After diagonalization, the matrix (eq. 4) can be converted to a form

$$\overleftrightarrow{Q} = \begin{bmatrix} \lambda_x^2 & 0 & 0 \\ 0 & \lambda_y^2 & 0 \\ 0 & 0 & \lambda_z^2 \end{bmatrix} \quad (6)$$

where $\lambda_x^2 + \lambda_y^2 + \lambda_z^2 = R_g^2$.

4. The shape of the polymer is characterized using the relative shape anisotropy

$$\kappa^2 = \frac{3}{2} \frac{\lambda_x^4 + \lambda_y^4 + \lambda_z^4}{(\lambda_x^2 + \lambda_y^2 + \lambda_z^2)^2} - \frac{1}{2} \quad (7)$$

$\kappa^2$ is bound between 0 and 1. It is equal to 0 only if all the points are spherically symmetric and is equal to 1 only if all points lie on a line.

5. The length of the sides of the imaginary cuboid enclosing the polymer:

$$\begin{aligned} SL_x &= \max(x_i) - \min(x_i) \\ SL_y &= \max(y_i) - \min(y_i) \\ SL_z &= \max(z_i) - \min(z_i) \end{aligned} \quad (8)$$

describes the volume occupied by the polymer and thus determines its packing density. Here, $x_i, y_i, z_i$ are the components of the bead positions $\vec{r}_i$.

The second class of descriptors we use are learnt descriptors. In order to obtain these feature vectors, we pass the *(N,3)* position vectors through an encoder model (part of an autoencoder) which outputs a *k*-dimensional feature vector [15]. While it is not possible to fully understand what these features represent (like in the case of calculated descriptors), they are useful patterns that the model has learnt and will help in the final prediction. Further, we do cross-validation to find out the number of such features needed.

### 2.2. Prediction Model

Once we have both kinds of descriptors, we can concatenate them and pass them through the ML model to predict the "probability of occurrence". In order to train this model, we also require training data which consists of the fingerprints (input data) and their ground truth probability of occurrence (labels). Since the ground truth probability of occurrence is not known, we cannot directly save it from the simulation run. To this end, we come up with a novel approach of approximating the probability of occurrence of a configuration at equilibrium. We make use of Internal Energy (U) as a surrogate for calculating ground truth labels. Figure 1 shows the typical profile of U against the number of steps of MC simulation. $U_{max}$ is the maximum value of U and $U_{equilibrium}$ is the "equilibrium" value defined as the equilibrium-average of U.

Kumar, I. and Jha, P. K.: *Preprint submitted to Elsevier*

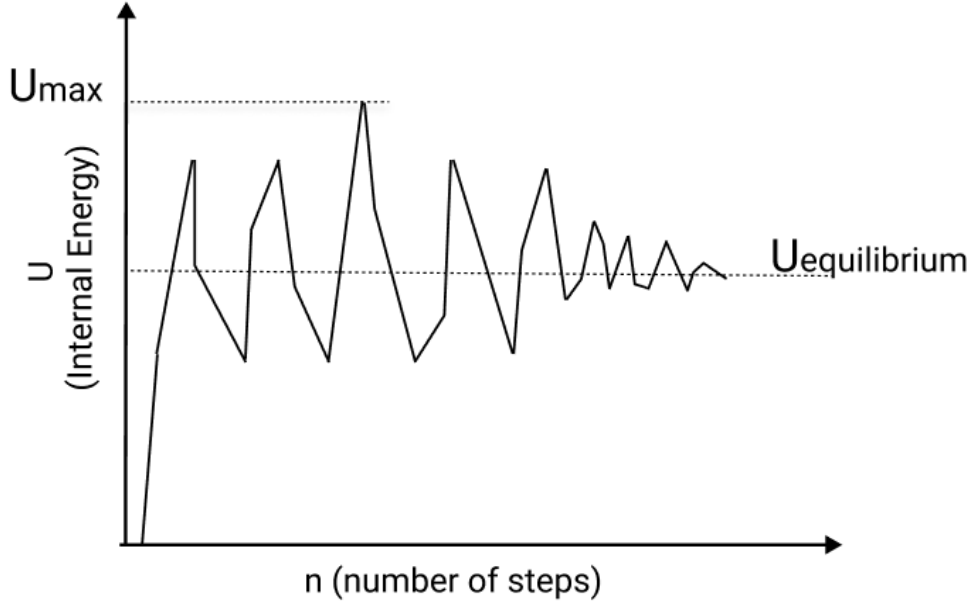

**Figure 1:** Plot of Internal Energy vs n. Values of $U_{max}$, $U_{equilibrium}$ can be found from it

The probability of occurrence of a configuration at equilibrium is defined as

$$P(U) = 1 - \frac{h(U)}{h_{max}} \qquad (9)$$

where $h(U)$ is the absolute difference from the equilibrium position

$$h(U) = |U - U_{equilibrium}| \qquad (10)$$

and $h_{max} = |U_{max} - U_{equilibrium}|$. This is equivalent to

$$P(U) = 1 - \frac{|U - U_{equilibrium}|}{|U_{max} - U_{equilibrium}|} \qquad (11)$$

Our approach ensures that $P(U) \to 0$ when $U \to U_{max}$ and $P(U) \to 1$ when $U \to U_{equilibrium}$. Note that the probabilities $P(U)$ should be interpreted as the relative probabilities of configurations not the absolute probabilities, as the latter should also satisfy the condition of normalization $\int P(U)dU = 1$

Once we obtain the labels $P(U)$ through the simulation run, our model can be trained. The architecture of the model consists of a fully connected neural network, which has a sigmoid nonlinearity since the output are probabilities and sigmoid ensures the range of the output is constrained between (0,1).

### 2.3. Details and Constraints of the Simulation process

We restrict this work to homopolymers (polymers consisting of only 1 type of repeat unit). However, our model can be easily extended to block copolymers also. Also, nonbonded interactions (e.g., electrostatic, van der Waals interactions, etc.) are not considered. The polymer chains are monodispersed with constant $N = 100$ (except for verification simulations in section 2.4). The first bead is initialised at origin, i.e., $\vec{r}_0 \equiv (0, 0, 0)$. For subsequent beads, the bond vectors $\vec{b}_i = \vec{r}_i - \vec{r}_{i-1}$ are randomly chosen with a uniform distribution in the range [-0.5, 0.5].

MC simulations consists of large number of MC steps involving trial displacements of a randomly chosen bead with maximum step size $\Delta R$ in a random direction. The displacements are made by picking a random bead and attempting a displacement on an imaginary sphere of fixed radius $\delta R \leq \Delta R$ around the current position. The maximum step size $\Delta R$ is a tunable parameter. It should neither be too small nor too large. If the $\Delta R$ is too small, it results in many moves being accepted, but the system advances slowly in configuration space. Many displacements are thus needed to achieve equilibration. On the other hand, if the $\Delta R$ is too large, many moves will be rejected and the equilibration process is again slow. We keep the step size as dynamic, i.e. it is increased when the acceptance percentage crosses a threshold, likewise decreases if it goes below a threshold. A running average of the acceptance percentage was kept for intervals of 400, 000 steps and we found that keeping the threshold as 68% worked well. MC simulation are conducted for $10^8$ MC steps. In order to ensure that the data points are decorrelated with each other the sampling is done with a frequency of $10^6$, i.e., states are saved after every $10^6$ steps.

The conventional Metropolis MC algorithm has an inherent bias towards generating high equilibrium probability configurations. The issue with this is, we will a strong imbalance in the dataset since we have hardly any configurations with low probability. This results in the prediction model not being able to learn that part of the dataset well. Thus, in order to generate low probability samples we make use of random sampling. We make a copy of our polymer state and apply random moves with a larger step size. If the final internal energy is less than a fixed $U_{max}$ then we accept it and save it as part of the dataset.

During the simulation run, we store values of various geometric descriptors at regular intervals defined using a sampling frequency. Along with these, we also store the position vectors of all beads at that time step, which is used to learn more useful descriptors using an unsupervised autoencoder model. Combining both of these (calculated and learnt) descriptors, we create a fingerprint and use it for the task of measuring probability of occurrence at equilibrium. The ground truth values of these probabilities is calculated using a linear relationship involving the internal energies of the run and we use a fully connected neural network to predict these probabilities given the fingerprint.

### 2.4. Verification through Scaling Laws

To verify our MC simulation method, we make use of two ideal chain scaling laws from literature: $Re^2$ and $Rg^2$ scale linearly with increase in $N$ (number of beads). The values of $Re^2$ and $Rg^2$ are calculated for simulations run for systems containing $N = 20 - 100$ beads and plotted in log-log scale. As evident from Figure 2, MC simulations confirm the linear scaling of $Re^2$ and $Rg^2$ with $N$.

## 3. Results and Discussion

The prediction model was trained on a dataset of combined fingerprint (i.e., including learnt and calculated descriptors) as input and the ground truth probabilities as output. The architecture is a fully connected neural network that regresses the probability, using 43 (32 learnt + 11 calculated) dimensional input fingerprint. The training and validation losses were monitored and it was found that the loss decreased smoothly during the course of training. We used AdamW as the optimizer for better regularization.[16]

In order to validate the performance of the Prediction model, certain other metrics were also used on the test data. Apart from the BCE Loss, The Root Mean Square Error (RMSE), Mean Absolute Error (MAE) and $R^2$ was also used. The values of the same are summarized in Table 1.

In order to validate the findings, we also tested using only the calculated descriptors, and not using the learnt descriptors. The input becomes 11 dimensional in this case. As it can be seen from the second column of Table 1 the values of metrics are lesser than the model using both descriptors. Hence, we can conclude that learnt descriptors add value to the model. Subsequently, we tested using only the learnt descriptors, and not using the calculated descriptors. The input becomes 32 dimensional in this case. As it can be seen from third column of Table 1, the values of metrics in this case are also lesser than the model using both descriptors. Hence we can conclude that calculated descriptors also add value to the model. From these tests, we find that the $R^2$ shows a value of > 0.9 in both calculated and learnt descriptor case. Therefore, if one is not interested in using the learnt descriptors in order to keep the model more explainable, then also a $R^2 \approx 0.92$ can be attained using the calculated descriptors alone. Further, we also verify the fact that both learnt and calculated descriptors provide useful information as removing either reduces the values of all metrics. The high $R^2$ value of 0.9949 for the model using both descriptors validates the idea that the configurational fingerprint adds value to the predictive capabilities of the model.

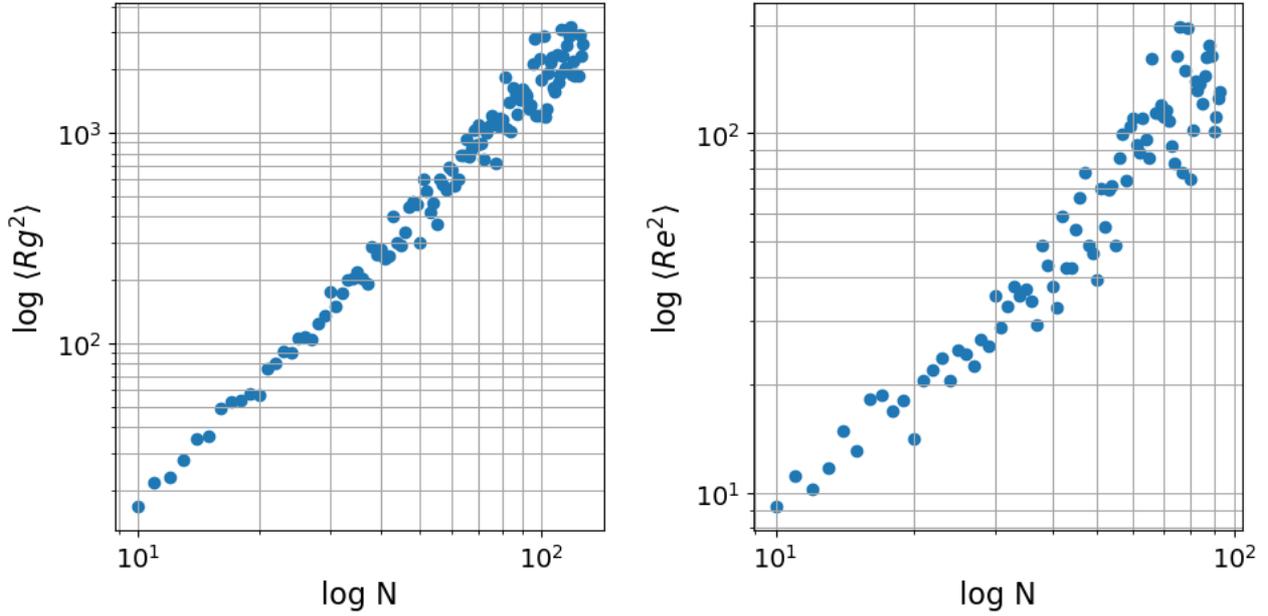

**Figure 2:** a) Scaling Laws for Rg² vs $N$ using a log-log plot. b) Scaling laws for Re² vs $N$ using a log-log plot.

**Table 1**
Results of metrics on test data using the fingerprint

| Metric | using both descriptors | using only calculated descriptors | using only learnt descriptors |
| --- | --- | --- | --- |
| Loss (BCE) | 0.2986 | 0.3477 | 0.4121 |
| RMSE | 0.01737 | 0.02939 | 0.03132 |
| MAE | 0.0126475 | 0.0149153 | 0.0161129 |
| R² | 0.9949 | 0.9249 | 0.9027 |

The residual curve is also plotted (Figure 3) to see differences between the actual and predicted values. The residual plot shows the ideal behaviour of being uniformly distributed around zero. Additionally, the residual plot shows that there are no correlations and most of the values lie between ±0.2. Anomalously high value at X≈75000 occurs because the position vector takes very large values in this case. It is a singular exceptional case when the MC simulation ends up taking very large steps, and therefore, due to lack of similar data, the model is unable to correctly predict it.

Configurational Polymer Fingerprints for Machine Learning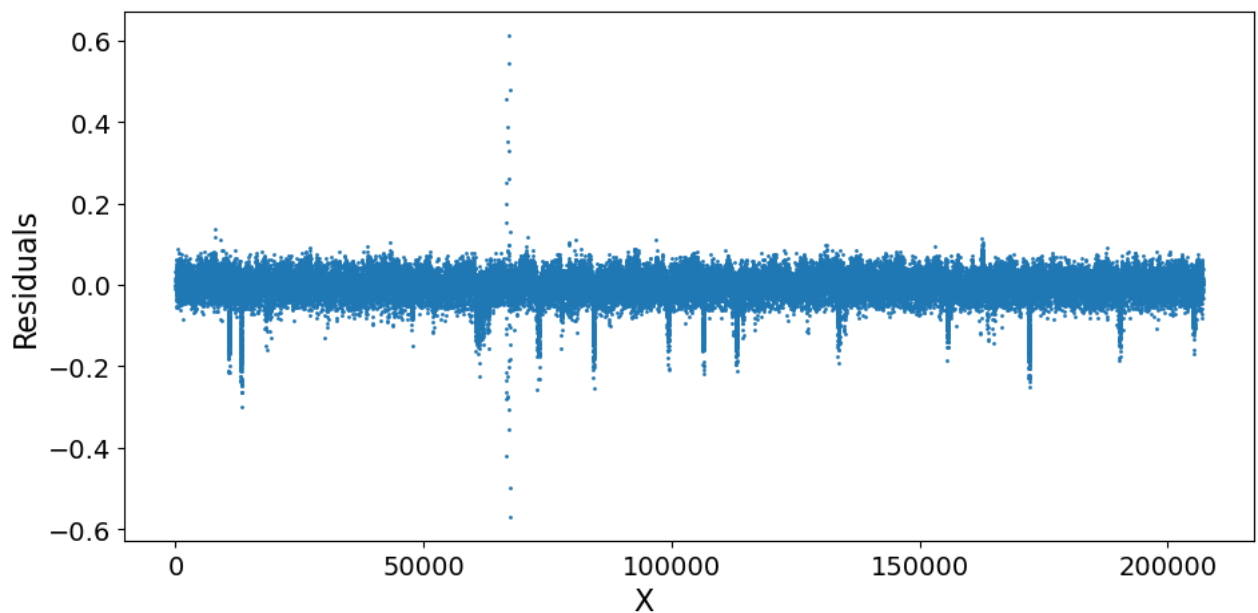

**Figure 3:** The residuals (difference of actual value and predicted value) plotted against X.

We have also tested the distribution of the input and output to see if they are similar and the low probability region is also being modelled. From the distributions in Figure 4, we observe that even the low probability regions are being accurately mapped by the model and it is not overfitting on the high probability zone.

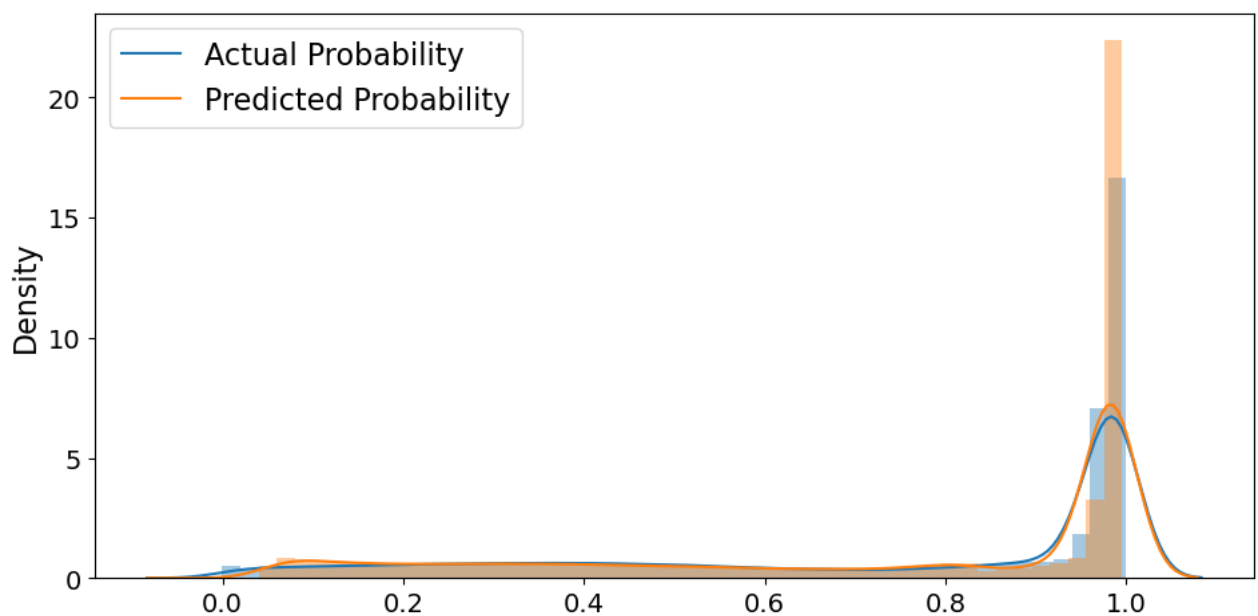

**Figure 4:** The difference of the output distributions between predicted and actual

## 4. Conclusions

From this work, we are able to show that a configurational fingerprint can be beneficial in predicting properties of a polymer using a generalised model of the polymer. We show that it performs well on the task of predicting the probability of occurrence at equilibrium. Further, we believe other tasks like QSAR/QSPR where existing methods only use atomic level information of the repeat unit, will also benefit from adding configurational fingerprint along. We also devise a method for creating ground truth labels of probability of occurrence at equilibrium, which can be used to train other models for predicting the same. Having an ML model predict probability of occurrence for given states also has large number of applications like finding ensemble averages of configuration dependent properties, calculating high probability/low probability configurations etc.

Several extensions of this study may be thought of. First, We only experimented using the basic bead spring model of linear polymers (did not consider branching), but the approach can easily be extended for branched polymers. Second, instead of only considering bonded interactions, we may also include non bonded interactions like Lennard Jones and Coulomb interactions, which may be required for modeling more realistic polymers. Third, the fully connected neural network can be replaced with a convolutional neural network so that it is spatially invariant.

## CRediT authorship contribution statement

**Ishan Kumar:** Conceptualization, Methodology, Software, Validation, Data Analysis, Writing - Original Draft. **Prateek K. Jha:** Conceptualization, Methodology, Supervision, Writing - Review & Editing.

The codes used in the manuscript is available at https://github.com/Ishan-Kumar2/configurational-polymerfingerp